\documentclass[prd,twocolumn,nofootinbib,amsmath,aps,floatfix,english,preprintnumbers,balancelastpage]{revtex4-1}
\usepackage{graphicx}
\usepackage{amsmath,bm}
\usepackage{longtable}
\usepackage[T1]{fontenc}
\usepackage{rotating}
\usepackage{float}
\floatstyle{plaintop}
\usepackage{tikz-feynman}
\usepackage{siunitx}

\usepackage{color}
\definecolor{naviBlue}{RGB}{0,0,128}
\usepackage[colorlinks  = true,
            linkcolor   = naviBlue,
            urlcolor    = naviBlue,
            citecolor   = naviBlue,
            anchorcolor = naviBlue]{hyperref}

\usepackage[normalem]{ulem}
\newcommand{\drop}[1]{}

\begin{document}

\title{Radiative neutrino masses and successful $\bf SU(5)$ unification}

\author{Christiane Klein}\email{christiane.klein@mpi-hd.mpg.de}
\author{Manfred Lindner}\email{manfred.lindner@mpi-hd.mpg.de}
\author{Stefan Vogl}\email{stefan.vogl@mpi-hd.mpg.de}
\affiliation{Max-Planck-Institut f\"ur Kernphysik,\\ Saupfercheckweg 1, D-69117 Heidelberg, Germany}

\begin{abstract}
Minimal $SU(5)$ Grand Unified models predict massless neutrinos and struggle to achieve gauge coupling unification compatible with the observed lower limit on the proton lifetime. Both of these issues can be resolved by embedding minimal radiative neutrino mass models into $SU(5)$. We systematically analyze the possible ways to realize radiative neutrino mass generation in  $SU(5)$ and provide a list of the minimal models. We find various models that have not been considered in the literature and demonstrate the compatibility of radiative neutrino masses with gauge coupling unification and proton decay for a new class of models with vector-like fermions. 
\end{abstract}

\maketitle

\section{Introduction}\label{sec:intro}
Grand-Unified-Theories (GUTs) are an attractive extension of the Standard Model (SM) and provide an elegant completion of the $SU(3)\times SU(2)\times U(1)$  gauge group of the Standard Model (SM). 
The simplest model for Grand Unification, the original $SU(5)$ model proposed by Georgi and Glashow \cite{Georgi:1974sy}, is known to exhibit a number of issues that seem to prevent a successful description of the world we observe. Among the most important of these are the inability to account for the observed  charged fermion masses,  the failure to achieve unification of the gauge couplings, rapid proton decay and the prediction of massless neutrinos. In order to address these issues extensions have been proposed which amend the model in various ways \cite{Georgi:1979df,Ellis:1979fg,Dimopoulos:1981yj,Sakai:1981gr,Dimopoulos:1981zb}. For a classic review of Grand Unified Theories see \cite{Langacker:1980js}, a more recent overview of GUT phenomenology can be found in \cite{Croon:2019kpe}.


One of the simplest extensions is a scalar $45$-plet. The $45$ contains an additional Standard-Model doublet, the interactions of which lead to additional contributions to the fermion masses and allow for the observed patters of the masses \cite{Georgi:1979df}. In addition it has been shown that this model also allows for unification at a scale high enough to suppress the proton decay rate  below the observed upper limit from Super-Kamiokande~\cite{Miura2016}. Nevertheless, the problem that neutrinos are predicted to be massless is not solved in this model.   

Further extensions of the field content are therefore unavoidable. As is well known from analyses of minimal Standard Model extensions neutrino masses can be generated at tree-level by the type-I, II and III seesaw mechanism \cite{Minkowski:1977sc,Mohapatra:1979ia,Yanagida:1979as,GellMann:1980vs,Schechter:1980gr,Mohapatra:1980yp,Foot:1988aq}. The required fields, i.e. complete SM-singlet fermions, scalar or fermionic $SU(2)_L$ triplets, can be embedded in $SU(5)$ theories.
However, it is also possible that the neutrino masses only arise at loop-level, see for example \cite{Zee:1980ai,Zee:1985id,Babu:1988ki} for early work in this direction and \cite{Cai:2017jrq} for a recent review of radiative neutrino mass models. Given that the minimal renormalizable $SU(5)$ does not contain the fields required for tree-level neutrino masses there is no preference for extensions that allow for seesaw scenarios over those with radiative mass generation.
In addition, the minimality of the theory, which is often a key guiding principle in the assessment of radiative neutrino mass models, could appear in a different light when seen from a UV-perspective instead of bottom-up. It is therefore of great interest to investigate all options for radiative neutrino masses in a GUT. Some general considerations concerning gauge coupling unification in radiative neutrinos masses can for example be found in \cite{Hagedorn:2016dze}.  The embedding of radiative neutrino masses into GUTs  in general and minimal renormalisable $SU(5)$ ($MRSU(5)$) in particular is not a completely new question, see \cite{Perez:2016qbo} for an example of a $SU(5)$-symmetric realization of the Zee-model or \cite{Saad:2019vjo} for a two-loop neutrino mass model.  However, a systematic study of possible UV-completions has not been undertaken so far. We will fill this gap in this work. 
Interestingly, we find that the additional multiplets required by the mass generation mechanism also provide the fields that allow for successful unification.

The structure of the paper is as follows. First, we comment on different ways to generate neutrino masses and analyze which minimal model for radiative neutrino masses can be realized in $SU(5)$ GUTs in Sec.~\ref{sec:RadMasses}. The one-loop contribution of models with new fermions is analyzed in Sec.~\ref{sec:RadMasses2}. Next, we introduce the conditions for successful unification of the gauge couplings and phenomenologically viable proton lifetime in Sec.~\ref{sec:Unification} before discussing the phenomenology of the viable models in more detail in Sec.~\ref{sec:Pheno}. Finally, we summarize our conclusions in Sec.~\ref{sec:summary}.

\section{$\bf SU(5)$-Embedding of radiative neutrino masses}\label{sec:RadMasses}

As mentioned previously, the minimal field content of a renormalizable $SU(5)$ theory that can accommodate the observed charged fermion masses and achieve unification in agreement with current proton decay limits predicts vanishing neutrino masses \cite{Perez:2016qbo}. Since at least two neutrinos must have non-zero masses in order to explain the observed neutrino oscillations an extension of the theory is unavoidable. In order to systematically track down all possible ways of incorporating radiative neutrino masses in a GUT framework, we start from the classification of radiative neutrino masses in terms of the minimal required field content presented in \cite{Klein:2019iws}. In the following, we use the notation introduced therein for the different realizations of radiative masses. By mapping the gauge quantum numbers of the particles found in the classification at the electroweak scale to the low-energy decomposition of the $SU(5)$-multiplets we can identify all ways of completing simple radiative neutrino mass models. In this study we limit ourself to moderately sized $SU(5)$-multiplets and do not investigate representations larger than $75$. A list of the decomposition of the relevant $SU(5)$ multiplets up to $75$ and a more general discussion about the group theoretical decomposition  can be found in \cite{Slansky1981}. 
In general, all models require at least two new SM multiplets which can be either scalars or fermions. We will list all models with a "minimal" field content that lead to neutrino masses. However, we will restrict ourselves to the models in which the standard seesaw-scenario cannot be realized. Similarly, we include models that predict unacceptable ratios of the charged fermion masses at the renormalizable level in the tables but disregard them  in the ensuing discussion. Regarding notation, we use $\phi_i$ to denote scalar fields and $\psi_i$ for fermions, where $i$ indicates the dimension of the $SU(5)$-representation. 
\\
{\bf Scalar models:} The models that only require the addition of new scalar representations  are listed in Tab.~\ref{tab:S}.

\setlength{\LTcapwidth}{3in}
\begin{longtable}[!tb]{ c  p{3.5cm}  p{3.5cm}}
\hline
Model & Scalar content & Neutrino mass model\\ \hline \endhead 
S1 & $\phi_5$, $\phi_{24}$,  $\phi_{10}$ &  cA2,(cA1), (A4)\\ \hline
S2 & $\phi_5$, $\phi_{24}$,   $\phi_{15}$ & seesaw type II, (cA5)\\ \hline
S3 & $\phi_5$, $\phi_{24}$, $\phi_{10}$,   $\phi_{40}$ & cA2,(cA1), (A4),(A2) \\ \hline 
S4 & $\phi_5$, $\phi_{24}$, $\phi_{10}$, $\phi_{50}$ & cA2, (A1),(cA1), (A4)\\ \hline
S5 & $\phi_5$, $\phi_{24}$, $\phi_{10}$, $\phi_{70}^1$, $\phi_{70}^2$ & cA2,(A5),(cA1), (A4)\\ \hline \hline
S6 & $\phi_5$, $\phi_{24}$, $\phi_{45}$,  $\phi_{10}$ & A0, cA2, cA4, (A4), (cA1), (cA3), (cA6)\\ \hline
S7 & $\phi_5$, $\phi_{24}$, $\phi_{45}$,  $\phi_{15}$ & seesaw type II, (cA5), (cA7)\\ \hline
S8 & $\phi_5$, $\phi_{24}$, $\phi_{45}$, $\phi_{40}$, $\phi_{50}$ & A3 \\ \hline \\
\caption{Neutrino mass in SU(5) GUT, extending the scalar sector only. Mass models in brackets either predict nonviable mass relations, are higher loop order or of higher effective operator dimension.}
\label{tab:S}
\end{longtable}

We find eight different ways to embed this class of radiative neutrino masses in $SU(5)$. The models S1-S5 do not contain the $45$ and require additional new physics for the generations of the observed charged fermion masses. In addition, we find two models, S2 and S7, that contain the scalar $SU(2)$-triplet  required by the type-II seesaw.  It would be unexpected if the radiative contributions are responsible for the observed neutrino masses and we do not consider these models further. This leaves the models S6 and S8 which fulfill all the quality criteria we have defined for an attractive  $SU(5)$-symmetric radiative neutrino mass model. Both these model have been discussed in the literature before. Model S6 extends the scalar content of $MRSU(5)$ by a $10$. This allows for a $SU(5)$ realization of the classic Zee model and  neutrino masses  are generated at one-loop. The phenomenology has been analyzed recently \cite{Perez:2016qbo} and it has been shown that this model allows for successful unification provided that some of the new scalars get a mass $\lesssim 100$~TeV. In addition, there are two other one-loop contribution to the neutrino mass, cA2 and cA4 in the notation of \cite{Klein:2019iws}. These contributions contain scalar leptoquarks and can also induce neutrino masses of the correct order of magnitude \cite{Dorsner:2017wwn}. In the second possible model, S8, the required addition consists of two scalar multiplets, a $40$ and $50$. This leads to neutrino masses at the two-loop level. Viable solutions for neutrino masses and unification  in the $SU(5)$-completion of S8 have been pointed out previously \cite{Saad:2019vjo}.
\\      
{\bf Mixed scalar and fermionic models:}
 Models requiring both new scalar and new fermionic representations  are listed in Tab.~\ref{tab:F}.
 
 \begin{longtable}[!tb]{ c  p{2.0cm}  p{2.0cm}  p{3cm} }
\hline
Model & Scalar content& Fermionic content & Neutrino mass model\\ \hline \endhead
F1 & $\phi_5$, and $\phi_{24}$ & $\psi_{\overline{5}}$, $\psi_{10}$,  $\psi_1$ & seesaw type I \\ \hline
F2 & $\phi_5$, and $\phi_{24}$ & $\psi_{\overline{5}}$, $\psi_{10}$,  $\psi_{24}$ &\parbox[c]{4.cm}{ seesaw type I+III,\\ (cB8), (cC1)} \\ \hline
F3 & $\phi_5$, $\phi_{24}$,  $\phi_{35}$ & $\psi_{\overline{5}}$, $\psi_{10}$, $\psi_{15}$,  $\psi_{\overline{15}}$ & B5 \\ \hline \hline
F4 & $\phi_5$, $\phi_{24}$,  $\phi_{45}$ & $\psi_{\overline{5}}$, $\psi_{10}$, $\psi_1$ & seesaw type I \\ \hline
F5 & $\phi_5$, $\phi_{24}$,  $\phi_{45}$ & $\psi_{\overline{5}}$, $\psi_{10}$,  $\psi_{24}$ & \parbox[c]{4cm}{seesaw type I+III, \\(cB8), (cB13), (cC1)} \\ \hline
F6 &  $\phi_5$, $\phi_{24}$, $\phi_{45}$ & $\psi_{\overline{5}}$, $\psi_{10}$, $\psi_{75}$ &\parbox[c]{4cm}{ seesaw type I, (cB8), \\(cB13), (cC1), (cC3)} \\ \hline 
F7 &  $\phi_5$, $\phi_{24}$, $\phi_{45}$, $\phi_{40}$ & $\psi_{\overline{5}}$, $\psi_{10}$, $\psi_{5}$, $\psi_{\overline{5}}$ & B3 \\ \hline
F8 &  $\phi_5$, $\phi_{24}$, $\phi_{45}$,  $\phi_{40}$ & $\psi_{\overline{5}}$, $\psi_{10}$, $\psi_{10}$,  $\psi_{\overline{10}}$ & B2 \\ \hline
F9 &  $\phi_5$, $\phi_{24}$, $\phi_{45}$, $\phi_{40}$ & $\psi_{\overline{5}}$, $\psi_{10}$, $\psi_{15}$,  $\psi_{\overline{15}}$ & B4 \\ \hline
F10 &  $\phi_5$, $\phi_{24}$, $\phi_{45}$, $\phi_{40}$ & $\psi_{\overline{5}}$, $\psi_{10}$, $\psi_{45}$, $\psi_{\overline{45}}$ & B3 \\ \hline  \\ 
\caption{Same as Tab.~\ref{tab:S} for models extending  the scalar and the fermionic sector. Models in which a change of the scalar sector alone leads to neutrino masses are omitted.}
\label{tab:F}
\end{longtable}
 
There are eight different ways to embed minimal radiative neutrino mass models with an additional fermion into $SU(5)$. For completeness, we also list the simplest possibilities for type-I seesaw. The first three listed possibilities do not have a scalar $45$ and, therefore, we omit them from our discussion. Three more models contain fermionic singlets and/or triplets and allow for the generation of neutrino masses by the type-I and III seesaw mechanism. This leaves four models, F7-F10, which deserve closer attention.  In F7 and F10 the neutrino mass is generated by two-loop diagrams while F8 and F9 already predict non-zero $m_\nu$ at one-loop order. To the best of our knowledge neither of these models has been discussed in the literature so far.  A GUT realization of neutrino masses based on diagrams with the same topology as F8 is discussed in \cite{Kumericki:2017sfc}. However, their field contend already allows for the generation of neutrino masses at one-loop without any extension of the fermion sector and, therefore, does not fulfill our minimality conditions.

{\bf Fermionic models:}
Finally, we would like to comment on the possibility of models without additional scalars.
It has been shown in \cite{Klein:2019iws} that purely fermionic extensions of the SM that generate neutrino masses only at the loop-level require at least three new fermions, a triplet, 4-plet and 5-plet of $SU(2)_L$. In particular the $4$- and $5$-plet are hard to embed in a complete $SU(5)$ multiplet and we do not find any UV-completion of purely fermionic extensions which fulfills our limit on the size of the representation. Nevertheless, we would like to comment that the fields can in principle be embedded in $SU(5)$. One minimal possibility is to take  the triplet from in a $15$, the 4-plet from a $70$ and the 5-plet from a $200$.   

\section{One-loop masses with new fermions}\label{sec:RadMasses2}

\begin{figure}[!t]
\centering
\begin{tikzpicture}
\begin{feynman}
\vertex (a){\(\nu_L\)};
\vertex[right=of a](x);
\vertex[right=of x](y);
\vertex[right=of y](z);
\vertex[right=of z](b){\(\nu_L\)};
\vertex[above=of y](w){\(h_{1/2}^+\)};
\diagram*[small, horizontal=a to b, inline=(a)]{
(a)--[fermion](x)--[anti fermion, edge label'=\(\psi_L\), insertion=0.99](y)--[anti fermion, edge label'=\(\psi_R\)](z)--[anti fermion](b),
(x)--[scalar, quarter left](w)--[scalar, quarter left](z),
};
\end{feynman}
\end{tikzpicture}
\caption[Schematic diagram of the neutrino mass term in models F8 and F9.]{Schematic diagram of the neutrino mass term in models F8 and F9. While the single diagram leads to a divergent result, the sum of the contributions of $h_1^+$ and $h_2^+$ running in the loop is finite. The fermion $\psi$ is the charged singlet (1, 1,-1) from the vector-like 10-plet in model F8, or a component of the $SU(2)_L$ triplet (1,3,-1) in the vector-like 15-plet in model~F9.}
\label{fig:diag}
\end{figure}
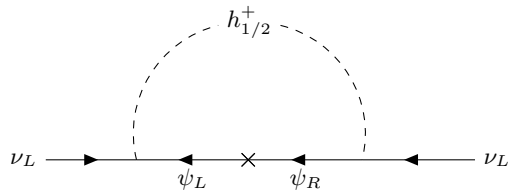

In order to get a better understanding of the physics in the one-loop models with new fermions a closer look at the expected neutrino masses is in order.
We start with model F8 (F9 leads to similar results). The part of the Lagrangian relevant for the neutrino mass is 
\begin{align}
\mathcal{L}_m &= \psi_{\bar{5}}^{SM}\psi_{10}^{SM}(Y_1^*\phi_5^*-\frac{1}{6}Y_2^*\phi_{45}^*) \nonumber \\ 
&+\psi_{\bar{5}}^{SM}\psi_{10} (Y_3^*\phi_5^*-\frac{1}{6}Y_4^*\phi_{45}^*) \nonumber \\
&+Y_5\psi_{\bar{5}}^{SM}\psi_{\bar{10}}\phi_{40}^*+ m_F \psi_{\bar{10}}\psi_{10}\nonumber \\ 
 &+ \lambda_1 \phi_5^2\phi_{45}\phi_{40} + \lambda_2 \phi_5 \phi_{45}^2 \phi_{40}\, \, .
\end{align}
When the two Higgs doublets in $\phi_5$ and $\phi_{45}$ develop a vacuum expectation value, the charged components mix according to 
\begin{align}
\begin{pmatrix} H_5^+ \\ H_{45}^+ \end{pmatrix} = \begin{pmatrix}
\cos \beta & \sin \beta \\ -\sin \beta & \cos \beta \end{pmatrix} 
\begin{pmatrix} G^+ \\ H^+ \end{pmatrix}\, \, ,
\end{align} 
where $G^+$ is the charged Goldstone boson absorbed by the gauge bosons, and
\begin{equation}
\tan \beta = -\frac{v_{45}}{v_5}\, ,
\end{equation}
with the two vacuum expectation values of the Higgs doublets in the 5-plet and 45-plet.
The charged Higgs field $H^+$ further mixes with the singly charged component $\Phi^+ \subset \Phi\sim (1,2,3/2)\subset \phi_{40}^*$:
\begin{align}
\begin{pmatrix} H^+ \\ \Phi^+ \end{pmatrix} =
\begin{pmatrix}\cos \theta & \sin \theta \\ -\sin \theta & \cos \theta \end{pmatrix} 
\begin{pmatrix} h_1^+ \\ h_2^+ \end{pmatrix}\, \, .
\end{align}
$\theta$ can be estimated as
\begin{align}
\tan 2\theta \approx \sin 2\theta = \frac{2\lambda v^2}{m_{h_1^+}^2-m_{h_2^+}^2}\; ,
\end{align}
and the effective coupling $\lambda$ is a combination of $\lambda_1$ and $\lambda_2$,
\begin{align}
\lambda &= \lambda_1 c_\beta(c^2_\beta -2s^2_\beta)+\lambda_2 s_\beta(s^2_\beta-2c^2_\beta)\, \, .
\end{align}
Here, we use the short notation $c_\beta = \cos \beta$, $s_\beta = \sin \beta$.
The neutrino mass generation in F8 is depicted in Fig.~(\ref{fig:diag}).
Neglecting the mixing of the fermions, we find that the neutrino mass is given by
\begin{align}
 m_\nu \approx  &\frac{Y_5Y_{new}^*}{2}s_{2\theta}
 \left(\mathcal{A}(m_F, m_{h_2^+})-\mathcal{A}(m_F, m_{h_1^+})\right)\\ 
 = &\frac{m_F v^2Y_5Y_{new}^*\lambda}{8\pi(m_{h_2^+}^2-m_{h_1^+}^2)(m_F^2-m_{h_1^+}^2)(m_F^2-m_{h_2^+}^2)}\nonumber \\
 &\times \left[m_F^2m_{h_2^+}^2\log\left(\frac{m_F^2}{m_{h_2^+}^2} \right)+m_F^2m_{h_1^+}^2\log \left( \frac{m_{h_1^+}^2}{m_F^2}\right)\right.  \nonumber \\
 &\; \; \left. +m_{h_1^+}^2m_{h_2^+}^2\log \left(\frac{m_{h_2^+}^2}{m_{h_1^+}^2}\right)\right]\; ,
 \label{eq:numass}
\end{align}
where we have introduced the shorthand notation $s_{2\theta}=\sin 2 \theta$. The effective Yukawa coupling reads
\begin{align}
Y_{new} &= Y_3 s_\beta -\frac{1}{6} Y_4 c_\beta
\end{align}
This result is in agreement with the results for a similar one-loop model obtained in \cite{Brdar:2013iea}, and, in the appropriate limit, with the results for the Zee-model \cite{Zee:1980ai}.

While we expect $h_1^+$, which is to good approximation the charged component of a low energy 2HDM, to be comparatively light, the other masses can be much heavier.
In the limit where $m_{h_2^+}$ is the largest of the masses appearing, the scale of the neutrino mass is proportional to 
\begin{align}
m_\nu \propto \frac{m_F v^2}{m_{h_2^+}^2}\, .
\end{align}
If the mass $m_F$ of the new fermion is the largest scale appearing in eq.~(\ref{eq:numass}), the dominant contribution to the neutrino mass is proportional to 
\begin{align}
m_\nu \propto \frac{v^2}{m_F}\, \, .
\end{align}
Hence, depending on the regime, the neutrino mass leads to different bounds on the masses $h_2^+$ and the new fermions $\psi$.  The topology of the loop diagram in model F9 is similar.  Therefore, we expect qualitatively and quantitatively similar results.

\section{Unification and proton decay}\label{sec:Unification}
In the following, we examine whether successful gauge coupling unification can be achieved within the new class of models featuring a mixed scalar-fermionic extension of the field contents. Since the scalar field content is similar this discussion applies to  F7-F10. 

 At one-loop, gauge coupling unification is described by~\cite{Giveon:1991zm} 
\begin{align}
\label{eq:Guteq}
\alpha_{GUT}^{-1}= \alpha_i^{-1}(M_Z)& - \frac{b_i}{2\pi}\ln \left(\frac{M_{GUT}}{M_Z}\right)\, ,
\end{align}
where $i=1,2,3$ runs over the SM gauge couplings and the one-loop $\beta$-functions $b_i$ 
\begin{align}
\label{eq:running}
b_i & = b_i^{SM}+ \Delta b_{i,k} r_k\, \, ,
\end{align}
can be decomposed into the SM one-loop $\beta$-functions
\begin{align}
b_1^{SM}&= \frac{41}{10}\, , & b_2^{SM}&=-\frac{19}{6}\, , & b_3^{SM}&= -7 \, ,
\end{align}
and the contributions $\Delta b_{i,k}$ of the new particles  that need to be  rescaled by
\begin{align}
r_k &= \frac{\ln (M_{GUT}/M_k)}{\ln (M_{GUT}/M_Z)} 
\end{align}
to account for the scale $M_k$ above which they contribute to the running.
An efficient way to check for unification is to rewrite (\ref{eq:Guteq}) in terms of 
\begin{align}
B_{ij} = b_i-b_j\, ,
\end{align}
 and the SM couplings at the electroweak scale, i.e. the finestructure constant $\alpha$, the strong coupling $\alpha_S$, and the electroweak mixing angle $\sin^2 \theta_W$. This leads to the equations \cite{Giveon:1991zm} 
\begin{align}
\ln \left(\frac{M_{GUT}}{M_Z} \right) = \frac{16\pi}{5\alpha (M_Z)}\frac{(3/8-\sin^2 \theta_W (M_Z))}{B_{12}} 
\label{eq:Gutscale}
\end{align}
and
\begin{align}
\frac{B_{23}}{B_{12}} &= \frac{5}{8} \left(\frac{\sin^2\theta_W (M_Z)-\alpha (M_Z) \alpha^{-1}_S(M_Z)}{3/8-sin^2\theta_W(M_Z)} \right) \nonumber \\ 
&= 0.719 \pm 0.005\, \, .
\label{eq:Unicond}
\end{align}
As is well-known, the SM alone does not lead to unification. This is reflected in the SM prediction $B_{23}/B_{12}\approx 0.5$, which is too small.

Another constrain on $B_{12}$ can be derived from the requirement that the proton lifetime should be sufficiently large, since the new gauge boson in $SU(5)$ unified theories mediate proton decay \cite{Georgi:1974sy}. In the absence of additional contributions to the decay, the lifetime of the proton can be estimated as \cite{Nath2006} 
\begin{align}
\tau_P \approx \frac{M_{GUT}^4}{\alpha_{GUT}^2 M_P^5} \, \, .
\end{align}
With this result and Eq.~(\ref{eq:Gutscale}), the current limit on the lifetime \cite{Miura2016}
\begin{align}
\tau_P \gtrsim \SI{1.6e34}{y}
\end{align} 
can be translated into a limit on the parameter $B_{12}$:
\begin{align}
\label{eq:Pcond}
B_{12} \leq 5.8 \, ,
\end{align}
the exact value depending on $\alpha_{GUT}$. Here we use a conservative estimate of $\alpha_{GUT} = 0.05$.
The SM value is $B_{12}=109/15$. In a successful unification scenario, $B_{12}$ must be lowered, so the unification scale is in agreement with proton decay bounds.
Therefore, beyond the SM fields, for instance the content of new multiplets required by neutrino mass generation, need to play a role in unification.  We focus on models F7-F10 in the following. 

Assuming that there is no mass splitting in the new fermionic multiplets, the fermions influence the running of all couplings  identically. Therefore, they do not help with unification and have no impact on the unification scale $M_{GUT}$. However, they can not be neglected completely since they have an impact on the value of the coupling $\alpha_{GUT}$ and might lead to a Landau-pole below the GUT scale.

In order to satisfy Eq.~(\ref{eq:Pcond}), a negative contribution to $B_{12}$ is required. We consider only scalars that have a negative contribution. The masses of all other scalars are assumed to be around $M_{GUT}$, such that these fields do not alter the running of the gauge couplings.
One exception is the scalar multiplet $\Phi\sim (1,2,-3/2) \subset \phi_{40}$. It has a positive contribution to $B_{12}$, but is required for the neutrino mass generation. Thus we allow it to be lighter than $M_{GUT}$ and consider its effect on gauge coupling unification.
Additionally, the masses of scalars mediating proton decay are set to $M_{GUT}$. 

These selection criteria single out seven scalar multiplets from the field content of models F7-F10. The influence of these fields on the running of the gauge couplings is summarized in Tab.~(\ref{tab:dbs}) (see also Tab.~(II) in \cite{Saad:2019vjo}).
In order to test for successful unification in this setup, we generate a set of seven $r_k$-parameters at random and check the conditions (\ref{eq:Pcond}) and
\begin{align}
\label{eq:Unicond_app}
0.709\leq \frac{B_{23}}{B_{12}} \leq 0.729\, .
\end{align}
\setlength{\LTcapwidth}{3in}
{\renewcommand{\arraystretch}{1.5}
\begin{longtable}{|c|c|c|c|c|c|}
\hline
Field & $\Delta b_{1,k}$ & $\Delta b_{2,k}$ & $\Delta b_{3,k}$ & $\Delta B_{12,k}$ & $\Delta B_{23,k}$ \\ \hline \endhead
$\phi_{(1,3,0)} \in 24$ & 0 & $\frac{1}{3}$ & 0 & $-\frac{1}{3}$ & $\frac{1}{3}$ \\ \hline
\hline
$\phi_{(1,2,1/2)} \in 45$ & $\frac{1}{10}$ & $\frac{1}{6}$ & 0 & $-\frac{1}{15}$ & $ \frac{1}{6}$ \\ \hline
$\phi_{(8,2,1/2)} \in 45$ & $\frac{4}{5}$ & $\frac{4}{3}$ & $2$ & $-\frac{8}{15}$ & $-\frac{2}{3}$ \\ \hline \hline 
$\phi_{(1,2,-3/2)} \in 40$ & $\frac{9}{10}$ & $\frac{1}{6}$ & 0 & $\frac{11}{15}$ & $ \frac{1}{6}$ \\ \hline
$\phi_{(3,2,1/6)} \in 40$ & $\frac{1}{30}$ & $\frac{1}{2}$ & $\frac{1}{3}$ & $-\frac{7}{15}$ & $\frac{1}{6}$ \\ \hline 
$\phi_{(\overline{3},3,-2/3)} \in 40$ & $\frac{4}{5}$ & $2$ & $\frac{1}{2}$ & $-\frac{6}{5}$ & $\frac{3}{2}$ \\ \hline 
$\phi_{(6,2,1/6)} \in 40$ & $\frac{1}{15}$ & $1$ & $\frac{5}{3}$ & $-\frac{14}{15}$ & $-\frac{2}{3}$ \\ \hline 
\caption[Change of the running of the gauge couplings for scalars in (F7)-(F10).]{Change of the one-loop running of the gauge couplings and the differences $\Delta B_{ij, k}=\Delta b_{i,k}-\Delta b_{j,k}$ for the new scalar fields in the models (F7)-(F10).}
\label{tab:dbs}
\end{longtable}
}
If the conditions are not satisfied, the parameter set is discarded.
We restrict the parameter $r_2$ describing the mass of the second Higgs doublet to the range $\left[0.85, \, 0.95\right]$. 

In the model F8, the new fermionic SU(5) multiplets will automatically receive a mass splitting of the order of the electroweak scale due to their Yukawa interactions. To make sure that the fermions do not influence the running of the couplings, and to avoid other possible problems of light vector-like fermions such as flavor-violation~\cite{Ishiwata:2015cga}, we assume that they are heavy enough for the splitting to be negligible, $m_F \geq\SI{e5}{GeV}$. As a result, a limit on the scalar mass can be derived by considering the neutrino mass. For $\mathcal{O}(1)$ Yukawa couplings, viable neutrino masses can be achieved if the mass of the scalar is larger than $\SI{e11}{GeV}$.  Thus we restrict its r-value to $  r_7 \leq 0.55$. Additionally, an upper limit for the scalar mass can be inferred from the neutrino mass. In the regime where the fermion mass $m_F$ is larger than $m_{h_2^+}$, $m_F$ is determined by the neutrino mass and gives an upper bound for $m_{h_2^+}$ of $\SI{e15}{GeV}$ for $\mathcal{O}(1)$ Yukawa couplings. The same upper limit can also be derived from the regime where $m_{h_2^+}$ is the largest mass scale in the neutrino mass diagrams. In this regime, the factor $\frac{m_\psi}{m_{h_2^+}}$ can at most be one, in which case, for $\mathcal{O}(1)$ Yukawa couplings, the scalar mass is fixed by the  neutrino mass to be at most $\SI{e15}{GeV}$. This corresponds to a bound  $0.15 \leq r_7$.  Taking this constraint into account the given solutions to the unification problem can also lead to viable masses for all fermions, including the neutrinos.

\section{Phenomenology}\label{sec:Pheno}
The parameter scan reveals that gauge coupling unification in agreement with proton decay and neutrino mass bounds is possible in the model F8. There is even a large range of parameters for which this can be achieved. The following two benchmark point illustrate that range.

\begin{figure}
\centering
\includegraphics[width=0.4\textwidth]{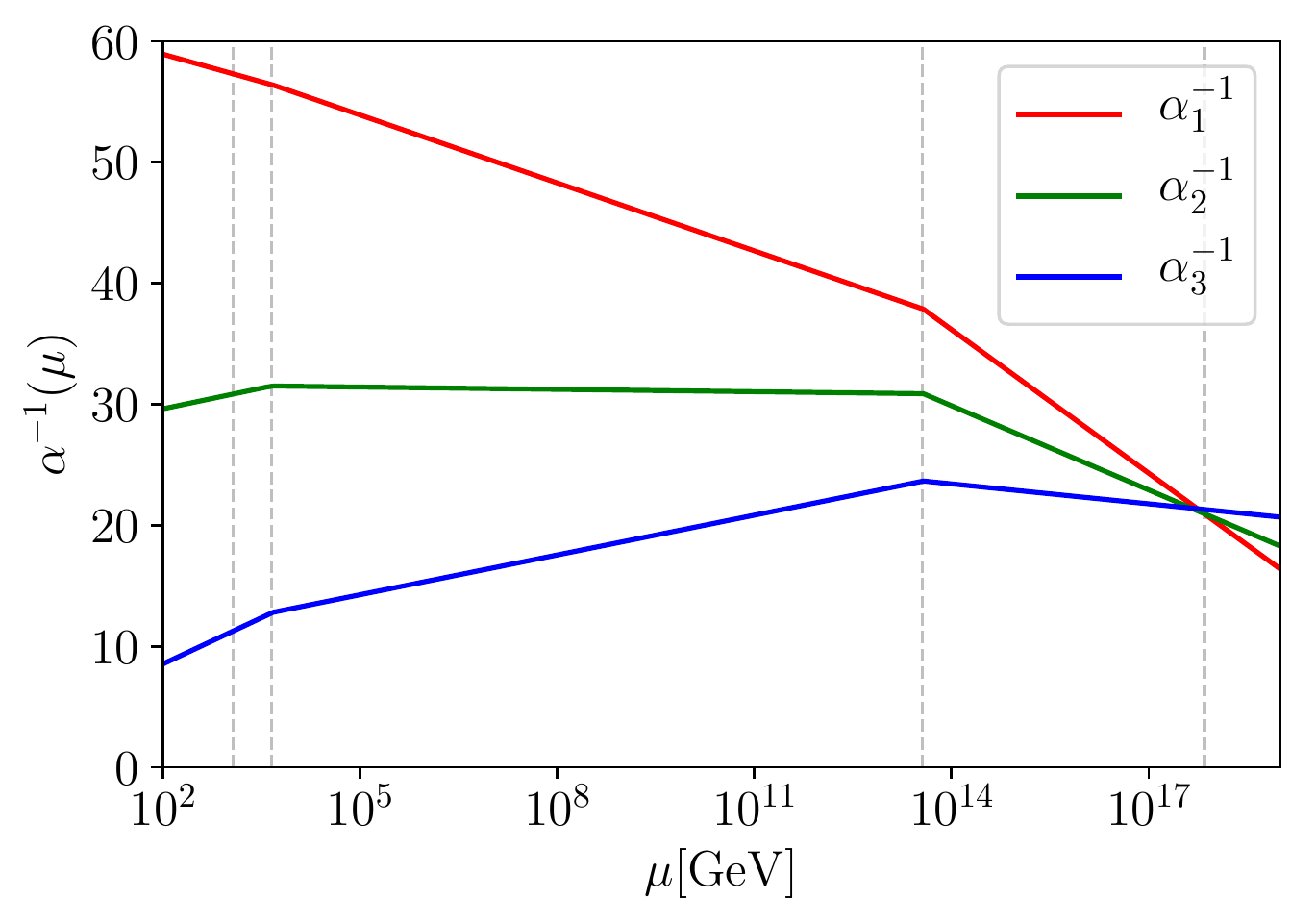}
\caption{Running of the gauge couplings for the benchmark point $m_{\phi_{(1,2,1/2)}} = \SI{1200}{GeV}$, $ m_{\phi_{(1,3,0)}} = m_{\phi_{(8,2,1/2)}}= m_{\phi_{(3,2,1/6)}} = m_{\phi_{(6,2,1/6)}} =\SI{4600}{GeV}$, $m_{\phi_{(\overline{3},3,-2/3)}} = m_{\phi_{(1,2,-3/2)}} = \SI{3.6e13}{GeV}$,  $m_F = \SI{3.6e13}{GeV}$, and  $ M_{GUT} = \SI{7e17}{GeV}$.}
\label{fig:r2}
\end{figure}

{\bf Large $\bf M_{GUT}$:} The largest GUT scale achievable while maintaining the correct neutrino  mass is about $\SI{e18}{GeV}$. One benchmark point with such a high unification scale is  $m_{\phi_{(1,2,1/2)}}=\SI{1200}{GeV}$ and $ m_{\phi_{(1,3,0)}}  = m_{\phi_{(8,2,1/2)}}= m_{\phi_{(3,2,1/6)}} = m_{\phi_{(6,2,1/6)}} =\SI{4600}{GeV}$. The two remaining scalars $\phi_{(\overline{3},3,-2/3)}$ and $\phi_{(1,2,-3/2)}$ reside at an intermediate scale $m_{\phi_{(\overline{3},3,-2/3)}} = m_{\phi_{(1,2,-3/2)}} = \SI{3.6e13}{GeV}$. The unification scale in this scenario is $ M_{GUT}= \SI{7e17}{GeV}$. This benchmark point features only a small number of scalar mass scales. In addition, the mass of the new fermions is taken to be equal to intermediate scalar mass scale. The running of the couplings in this case, including two generations of the new fermions from F8, is shown in Fig.~(\ref{fig:r2}).
\begin{figure}
\centering
\includegraphics[width=0.4\textwidth]{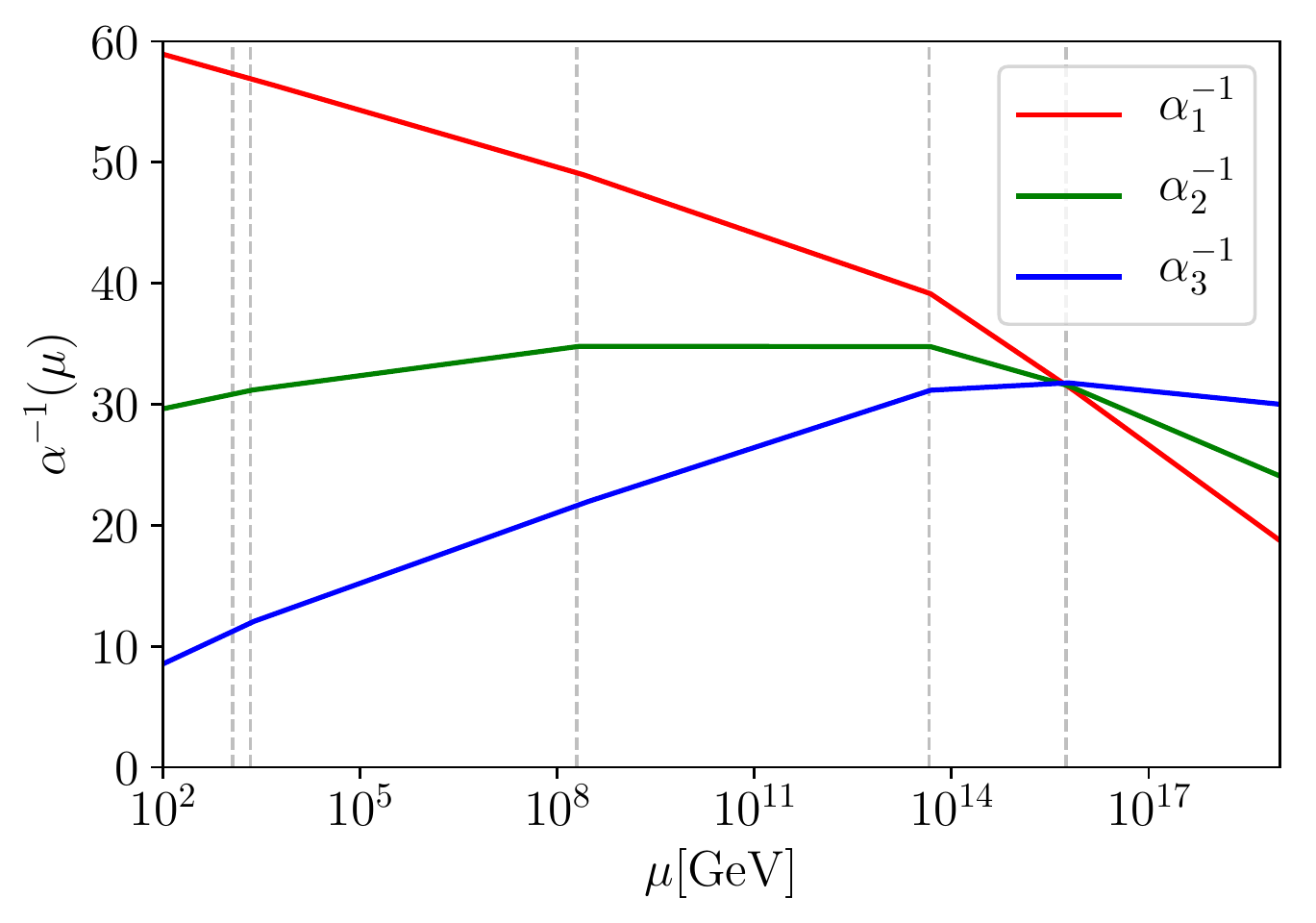}
\caption{Running of the gauge couplings for the benchmark point $m_{\phi_{(1,2,1/2)}} =\SI{1200}{GeV}$, $ m_{\phi_{(6,2,1/6)}} = \SI{2200}{GeV}$, $m_{\phi_{(\overline{3},3,-2/3)}} = \SI{2.0e8}{GeV}$, $ m_{\phi_{(1,2,-3/2)}} = \SI{4.7e13}{GeV}$,  $m_{\phi_{(1,3,0)}} = m_{\phi_{(8,2,1/2)}} = m_{\phi_{(3,2,1/6)}}= M_{GUT} = \SI{5.5e15}{GeV}$, and $m = \SI{4.6e13}{GeV}$.}
\label{fig:r3}
\end{figure}
To achieve a high unification scale, many of the new particles which push the scale up need to be light. Among them is also a scalar color octet $\phi_{(8,2,1/2)}$.
At this benchmark point, the mass of the scalar color octet is very close to the current bound from collider search of $m_{\phi_{(8,2,1/2)}}\geq \SI{4.2}{TeV}$ \cite{Sirunyan:2018xlo}, and might be excluded in the near future. 

{\bf Low number of light fields:} 
A realization with just four light scalars, the second Higgs doublet and the fields $\phi_{(6,2,1/6)}$, $\phi_{(\overline{3},3,-2/3)}$, and $\phi_{(1,2,-3/2)}$ from the 40-plet, is also feasible. $\phi_{(6,2,1/6)}$, $\phi_{(\overline{3},3,-2/3)}$ are responsible for successful unification. The second Higgs doublet is expected to be light, while the field $\phi_{(1,2,-3/2)}$ is required to be light by the neutrino masses. Additionally, the new fermions have mass below the GUT scale as well. One benchmark point with this minimal number of light fields consists of two fields at the TeV-scale, $m_{\phi_{(1,2,1/2)}} = \SI{1200}{GeV}$ and  $m_{\phi_{(6,2,1/6)}} = \SI{2200}{GeV}$. The two other scalars reside at intermediate mass scales. In contrast to the previous case, their masses cannot be chosen equal with such a small number of light fields, so $m_{\phi_{(\overline{3},3,-2/3)}} = \SI{2.0e8}{GeV}$ and $ m_{\phi_{(1,2,-3/2)}} = \SI{4.7e13}{GeV}$. The fermion mass scale is $m_\psi = \SI{4.6e13}{GeV}$, while all other fields have masses of the order of the GUT scale $m_{\phi_{(1,3,0)}} = m_{\phi_{(8,2,1/2)}} = m_{\phi_{(3,2,1/6)}}= M_{GUT} = \SI{5.5e15}{GeV}$. The running of the couplings in this case is illustrated in Fig.~(\ref{fig:r3}). 
Note that this benchmark point predicts a proton lifetime of $\tau_p \approx \SI{2.7e34}{y}$, which is well within the reach of the planned Hyper-Kamiokande experiment \cite{Abe:2018uyc}.

\section{Conclusion}\label{sec:summary} 

We have investigated how gauge coupling unification and radiative neutrino masses can be addressed in minimal $SU(5)$ models. As a first step, the embedding of minimal radiative neutrino mass models into $SU(5)$ is demonstrated. Extensions with scalars and fermions allow for neutrino mass at one or two-loop. Insisting on a renormalizable generation of the charged fermion masses leads to nine distinct models with radiative contributions to the neutrino mass. Three of these can be realized with a purely scalar extension of the field content but in one also the type-II seesaw mechanism is viable. Hence, we find only two models for genuine radiative neutrino masses in minimal scalar extensions of renormalizable $SU(5)$. Both possibilities have been considered in the literature and it has been shown that they allow for successful unification \cite{Perez:2016qbo,Saad:2019vjo}. 
Extending both the scalar and the fermionic sectors leads to the remaining six models. We again find that only four of these predict genuine loop-level neutrino masses while two also predict contributions from the type-I and type-III seesaw.  
We  analyze the one-loop neutrino masses in models with new fermions and demonstrate that they provide options for a phenomenologically viable unification of the gauge couplings. Consequently, two of the main short comings of minimal $SU(5)$ GUTs can be addressed  simultaneously.

\vspace{1cm}
\section*{Acknowledgements}

We thank Sebastian Ohmer for collaboration on previous work and helpful discussions regarding radiative neutrino masses.


\bibliography{bibliography}{}

\end{document}